\documentclass[aps,prd,onecolumn,eqsecnum,amsmath,nofootinbib,preprintnumbers]{revtex4}%
\newcounter{example}[section]

\usepackage{amsmath}
\usepackage{color,graphicx,float,subfigure}
\usepackage{amsfonts,amssymb,theorem,mathrsfs,times}
\usepackage{bm}
\usepackage{tensor}
\newcommand{\D}{{\rm d}}

{\theorembodyfont{\upshape}
	}
{\theorembodyfont{\upshape}
	}
{\theorembodyfont{\upshape}
	}
{\theorembodyfont{\upshape}
	}
{\theorembodyfont{\upshape}
	}
{\theorembodyfont{\upshape}
	}

\newcommand{\dalm}{\kern1pt\vbox{\hrule height 0.9pt\hbox{\vrule width
			0.9pt\hskip 2.5pt\vbox{\vskip 5.5pt}\hskip 3pt\vrule width
			0.3pt}\hrule height 0.3pt}\kern1pt}

\begin{document}
	
	\title{The shadow radius lower bound is more robust than the photon sphere radius lower bound in $n$-dimensional Einstein gravity}
	
	\author{Jiaqi Fu\footnote{e-mail
			address: 1491713073@qq.com}}
	\author{Yong Song\footnote{e-mail
			address: syong@cdut.edu.cn (corresponding author)}}
	
	\affiliation{
		College of Physics\\
		Chengdu University of Technology, Chengdu, Sichuan 610059, China}
	
	\date{\today}

	\begin{abstract}
		The photon sphere and the black hole shadow are two closely related geometric features of strong-field spacetimes: the photon sphere is a hypersurface foliated by circular null geodesics, while the shadow boundary is determined by the critical impact parameter for null-geodesic capture. For static, spherically symmetric, asymptotically flat black holes in $n$-dimensional ($n\ge4$) Einstein gravity with anisotropic matter, we derive lower bounds on the shadow radius and the photon sphere radius and compare their matter assumptions. Assuming that an auxiliary function $\Xi(r)$ satisfies $-\rho\le\Xi\le p_r$ and $(r^2\Xi)'\ge0$ between the horizon and the photon sphere under consideration, we establish $r_\gamma\ge[(n-1)/2]^{1/(n-3)}r_H$, where $r_H$ is the event horizon radius. Writing the shadow radius as $r_{\mathrm{sh}}=[\max_{r>r_H}U(r)]^{-1/2}$, where $U(r)$ is the null effective potential, we obtain $r_{\mathrm{sh}}/r_H\ge[(n-1)/2]^{1/(n-3)}\sqrt{(n-1)/(n-3)}$ using only $\rho\ge0$ and $\rho+p_r\ge0$, both implied by the weak energy condition (WEC). This shadow bound holds irrespective of the number of photon spheres and does not require the additional matter conditions used in the photon sphere lower-bound proofs. In this sense, the lower bound on the shadow radius is more robust than that on the photon sphere radius. A four-dimensional configuration satisfying the WEC has a single exterior photon sphere and obeys the shadow bound while violating the photon sphere lower-bound inequality, demonstrating this distinction even when both radii are associated with the same photon sphere. Both lower bounds are saturated by the Schwarzschild--Tangherlini solution. In Appendix~\ref{appendixA}, we also derive the mass-dependent upper bounds under the WEC, the strong energy condition, and suitable asymptotic decay assumptions.
	\end{abstract}
	
	\maketitle
	
	\section{Introduction}
	Black holes, as fundamental predictions of general relativity and key astrophysical objects, continue to be a central focus of theoretical and observational research. The horizon-scale images of black holes obtained by the Event Horizon Telescope (EHT)~\cite{EventHorizonTelescope:2019dse,EventHorizonTelescope:2022wkp,Vagnozzi:2022moj} have provided insight into the strong-field regime of gravity and motivated the use of black hole shadows as probes of spacetime geometry near compact objects.
	
	A central geometric feature associated with the apparent size of the shadow is the photon sphere---a hypersurface foliated by null circular geodesics~\cite{Claudel:2000yi}. Circular null orbits also play important roles in rotating black hole geometries, gravitational lensing, and the eikonal description of quasinormal modes~\cite{Bardeen:1972fi,Cardoso:2008bp,Hod:2012ax,Virbhadra:1999nm,Stefanov:2010xz}. As demonstrated in Ref.~\cite{Guo:2022ghl}, black hole spacetimes can possess multiple photon spheres, including stable ones between unstable ones. For an asymptotically flat, static, spherically symmetric black hole, the shadow radius seen by an asymptotic observer is given by the critical impact parameter for null-geodesic capture.
	
   For the four-dimensional Schwarzschild black hole, the photon sphere is located at $r_{\gamma}=3M$ and the shadow radius satisfies $r_{\mathrm{sh}}=3\sqrt{3}M$. Photon sphere bounds in more general black hole spacetimes have been studied from several complementary viewpoints. For static, spherically symmetric, asymptotically flat black holes with matter fields, including hairy black holes~\cite{Nunez:1996xv,Hod:2011aa}, Hod established the mass-dependent upper bound $r_{\gamma,\mathrm{in}}\le3M$ for the innermost photon sphere under suitable energy conditions~\cite{Hod:2013jhd}. The relation between the photon sphere and the critical impact parameter was further investigated by Cveti\v{c}, Gibbons, and Pope~\cite{Cvetic:2016bxi}, and a sequence of inequalities relating the horizon, photon sphere, shadow, and mass was subsequently proposed and analyzed~\cite{Lu:2019zxb,Yang:2019zcn}. In particular, for the outermost photon sphere, Yang and L\"u proved $r_{\gamma,\mathrm{out}}\le3M$ under suitable assumptions and showed that the required energy conditions can be weakened to the null energy condition outside the outermost photon sphere~\cite{Yang:2019zcn}. By comparison, lower bounds on the photon sphere radius generally require additional assumptions on the matter distribution. Hod proved $r_{\gamma,\mathrm{in}}\ge3r_H/2$ for the innermost photon sphere of spherically symmetric black holes under the dominant energy condition, a non-positive trace condition, and an additional monotonicity condition on the radial pressure~\cite{Hod:2020pim}. Yang and L\"u obtained the same lower bound for the outermost photon sphere under an auxiliary-function condition involving $\Xi(r)$, and exhibited configurations showing that the WEC by itself is not sufficient~\cite{Yang:2019zcn}. For the shadow radius, the WEC alone implies $r_{\mathrm{sh}}/r_H\ge3\sqrt{3}/2$~\cite{Hod:2026jtr}, while the combination of WEC and SEC, together with suitable asymptotic falloff, yields $r_{\mathrm{sh}}\le3\sqrt{3}M$~\cite{Yang:2019zcn}. All bounds are saturated by the vacuum Schwarzschild solution. Related photon sphere and shadow bounds, including extensions to rotating and higher-curvature black holes, have also been discussed in Refs.~\cite{Feng:2019zzn,Ma:2019ybz}.
	
	With growing interest in higher-dimensional theories of gravity~\cite{Singh:2017vfr,Amir:2017slq,Nozari:2023flq,Singh:2023ops,Singh:2024rnh,Nozari:2024jiz,Nozari:2026dkn}---motivated by string theory, brane-world scenarios, and the gauge/gravity duality---it is natural to ask whether analogous bounds exist for the photon sphere and shadow radius in $n$-dimensional ($n\geq4$) Einstein gravity. Song et al.~\cite{Song:2026cci} recently derived photon sphere bounds in higher dimensions. Their conditional lower-bound analysis uses pressure bounds $\rho\ge|p_r|,|p_t|$, a non-positive trace of the energy-momentum tensor, and an additional monotonicity assumption. Although termed the WEC in that reference, these pressure bounds constitute the dominant energy condition and are stronger than the WEC used here. These results motivate a systematic investigation of the assumptions underlying the photon sphere and shadow lower bounds, and of whether the distinction between their domains of validity persists in arbitrary dimension.
	
	In this work, we derive lower bounds on the shadow radius and the photon sphere radius in $n$-dimensional Einstein gravity under suitable conditions on the matter fields, and compare the assumptions required for the two bounds. We first define photon spheres and the shadow radius through the null effective potential. Under the auxiliary-function condition stated below, we extend the four-dimensional argument to arbitrary dimension and obtain \(r_\gamma\ge[(n-1)/2]^{1/(n-3)}r_H\) for every exterior photon sphere to which the condition applies. We also show that the WEC and a pressure-monotonicity condition up to the innermost photon sphere imply the same lower bound for all exterior photon spheres, without requiring a non-positive trace. We then derive the lower bound \(r_{\mathrm{sh}}/r_H\ge[(n-1)/2]^{1/(n-3)}\sqrt{(n-1)/(n-3)}\) using only $\rho\ge0$ and $\rho+p_r\ge0$. Thus, within the WEC class, the shadow lower bound holds without the additional matter conditions used in the photon sphere lower-bound proofs. This is the sense in which we describe the shadow lower bound as more robust; robustness here refers to its validity under weaker matter assumptions, rather than to dynamical stability. The distinction remains meaningful when the two radii are associated with the same photon sphere: $r_*$ is the areal radius of a photon sphere attaining the global maximum of $U$, whereas $r_{\mathrm{sh}}$ is its critical impact parameter. We demonstrate the distinction explicitly with a four-dimensional configuration satisfying the WEC and possessing a single exterior photon sphere: its areal radius violates the photon sphere lower-bound inequality, while its critical impact parameter still satisfies the shadow lower bound. Both lower bounds are saturated by the Schwarzschild--Tangherlini solution. Table~\ref{tab:bounds} summarizes the bounds and their sufficient assumptions. The corresponding upper bounds are derived in Appendix~\ref{appendixA} under the WEC, the SEC, and suitable asymptotic decay assumptions.
	
	\begin{table}[htbp]
		\centering
		\caption{Lower and upper bounds on the photon sphere radius and the shadow radius for static, spherically symmetric, asymptotically flat black holes with a regular horizon. Here $r_\gamma$ denotes photon sphere radius. The auxiliary-function condition in Eq.~(\ref{Xi}) is needed only on $[r_H,r_\gamma]$ for a particular photon sphere. The alternative lower-bound argument assumes the WEC throughout the exterior and non-increase of $|r^{n-1}p_r|$ only on $[r_H,r_{\gamma,\mathrm{in}}]$, assuming the innermost exterior photon sphere exists. The remaining matter conditions apply throughout the exterior; $M=\lim_{r\to\infty}m(r)$. All bounds are saturated by the Schwarzschild--Tangherlini solution.}
		\label{tab:bounds}
		\renewcommand{\arraystretch}{1.5}
		\begin{tabular}{|c|c|}
			\hline
			\textbf{Bound} & \textbf{Sufficient assumptions} \\
			\hline
			\( r_\gamma \geq \left(\frac{n-1}{2}\right)^{\frac{1}{n-3}} r_H \) & \begin{tabular}{c} Auxiliary-function condition \\ or \\ WEC + non-increase of $|r^{n-1}p_r|$ \end{tabular} \\
			\hline
			\( \frac{r_{\mathrm{sh}}}{r_H} \geq \left(\frac{n-1}{2}\right)^{\frac{1}{n-3}} \sqrt{\frac{n-1}{n-3}} \) & \begin{tabular}{c} $\rho\ge0$ and $\rho+p_r\ge0$ \\ (implied by WEC) \end{tabular} \\
			\hline
			\( r_\gamma \leq [(n-1)M]^{\frac{1}{n-3}} \) & WEC + SEC + asymptotic decay \\
			\hline
			\( r_{\mathrm{sh}} \leq \sqrt{\frac{n-1}{n-3}}[(n-1)M]^{\frac{1}{n-3}} \) & WEC + SEC + asymptotic decay \\
			\hline
		\end{tabular}
	\end{table}
	
	The paper is organized as follows. In Sec.~\ref{section2} we describe the spacetime and present the field equations. Section~\ref{lower-bounds} is devoted to the lower bounds. We first define photon spheres and the capture shadow, then derive the conditional lower bound on the photon sphere radius and the WEC lower bound on the shadow radius. We subsequently compare their matter assumptions, explain why the shadow lower bound is more robust, and verify the distinction using a four-dimensional matter example with a single exterior photon sphere. Section~\ref{section6} summarizes our results. The upper bounds are derived in Appendix~\ref{appendixA}. Throughout we use natural units with $G_n=c=1$.

	\section{Description of the higher-dimensional black hole system}\label{section2}
	
	Consider a static, spherically symmetric, asymptotically flat black hole configuration in $n$-dimensional spacetime. The $n$-dimensional Einstein-Hilbert action with matter fields is given by
	\begin{align}
		\label{s}
		S = \int d^n x \sqrt{-g} \left(\frac{R}{16\pi} + \mathcal{L}_M\right)\;,
	\end{align}
	where $R$ is the Ricci scalar, $g$ the metric determinant, and $\mathcal{L}_M$ the matter Lagrangian.
	
	The spacetime metric can be written in Schwarzschild-like coordinates as~\cite{Song:2026cci}
	\begin{align}
		\label{metric}
		ds^{2} = -e^{-2\delta(r)}\mu(r)\, dt^{2} + \mu(r)^{-1} dr^{2} + r^{2} d\Omega_{n-2}^{2}\;,
	\end{align}
	where the metric functions $\delta(r)$ and $\mu(r)$ depend only on the areal coordinate $r$, and
	\begin{align}
		\label{omega}
		d\Omega_{n-2}^{2} = d\theta_{1}^{2} + \sin^{2}\theta_{1} d\theta_{2}^{2} + \cdots + \left(\prod_{i=1}^{n-3}\sin^{2}\theta_i\right) d\theta_{n-2}^{2}
	\end{align}
	represents the line element of the unit $(n-2)$-sphere.
	
	For the regular outer event horizon at $r = r_H$, we impose
	\begin{align}
		\label{mu(rH)}
		\mu(r_{H}) = 0  \quad  \text{with} \quad  \mu^{\prime}(r_{H}) \geq 0 \;.
	\end{align}
	We assume $\mu(r)>0$ for $r>r_H$, finite $\delta(r_H)$, and a smooth static exterior. The stress tensor is required to be regular in a freely falling frame at the horizon. Asymptotic flatness imposes
	\begin{align}
		\label{infty}
		\mu(r \rightarrow \infty) \rightarrow 1 \quad \text{and} \quad \delta(r \rightarrow \infty) \rightarrow 0 \;.
	\end{align}
	Here we do not assume $\delta(r)= 0$, so our results are also applicable to hairy black hole configurations~\cite{Volkov:1998cc,Volkov:2016ehx}.
	
	The energy-momentum tensor of the matter field is diagonal in the static orthonormal frame,
	\begin{align}
		\label{Tmunu}
		T^{\mu}{}_{\nu} = \operatorname{diag}\bigl(-\rho(r), p_{r}(r), \underbrace{p_{t}(r), p_{t}(r), \ldots, p_{t}(r)}_{n-2\ \text{terms}}\bigr)\;,
	\end{align}
	where $\rho$, $p_r$, and $p_t$ are the energy density, radial pressure, and tangential pressure, respectively. Spherical symmetry requires the $(n-2)$ tangential pressure components to be identical, while $p_r$ and $p_t$ may differ.
	
	The Einstein equations $G^{\mu}{}_{\nu}= 8\pi T^{\mu}{}_{\nu}$ yield
	\begin{align}
		\label{mu'}
		\mu^{\prime} &= \frac{(n-3)(1-\mu)}{r} - \frac{16\pi r\rho}{n-2}\;,\\
		\label{delta'}
		\delta^{\prime} &= -\frac{8\pi r\left(\rho+p_{r}\right)}{(n-2)\mu}\;,
	\end{align}
	where a prime denotes differentiation with respect to $r$. The horizon condition in Eq.~(\ref{mu(rH)}), together with Eq.~(\ref{mu'}) and regularity of the stress tensor at the horizon, gives
	\begin{align}
		\label{rho}
		\rho(r_{H}) \leq \frac{(n-3)(n-2)}{16\pi r_{H}^{2}}\;, \qquad p_{r}(r_{H}) = -\rho(r_{H})\;.
	\end{align}
	The geometrically normalized mass function $m(r)$ is defined by
	\begin{align}
		\label{m(r)}
		m(r) = \frac{1}{2} r_H^{n-3} + \frac{8\pi}{n-2} \int_{r_H}^r x^{n-2} \rho(x)\, dx\;,
	\end{align}
	where $\frac{1}{2} r_H^{n-3}$ is the horizon value $m(r_H)$. We denote the finite asymptotic mass parameter by $M\equiv m(r\to\infty)$. The physical ADM mass is related to this geometric normalization by
	\begin{align}
		\label{ADM-normalization}
		M_{\mathrm{ADM}}=\frac{(n-2)\Omega_{n-2}}{8\pi}M,\qquad
		\Omega_{n-2}=\frac{2\pi^{(n-1)/2}}{\Gamma[(n-1)/2]}\;.
	\end{align}
	From Eqs.~(\ref{mu'}) and (\ref{m(r)}), the relation between $\mu$ and $m(r)$ in $n$-dimensional spacetime takes the form
	\begin{align}
		\label{mu(mr)}
		\mu(r) = 1 - \frac{2 m(r)}{r^{n-3}} \;.
	\end{align}
	
	Substituting Eqs.~(\ref{mu'}) and (\ref{delta'}) into the energy-momentum conservation equation $\nabla_\mu T^{\mu}{}_{r} = 0$ yields
	\begin{align}
		\label{p'}
		p_r^{\prime} = \frac{2 T + (n-1)\rho - (n+1) p_r}{2 r} - \left(\frac{n-3}{2 r} + \frac{8\pi r p_r}{n-2}\right)\frac{\rho+p_r}{\mu}\;,
	\end{align}
	where
	\begin{align}
		\label{trace}
		T = -\rho + p_r + (n-2) p_t
	\end{align}
	is the trace of the energy-momentum tensor $T^\mu{}_{\nu}$.
	
	In the subsequent analysis we will invoke the weak energy condition (WEC) and the strong energy condition (SEC). For the stress tensor in Eq.~(\ref{Tmunu}), these conditions take the form~\cite{Kontou:2020bta}
	\begin{itemize}
		\item[(1)] Weak Energy Condition (WEC): The energy density is non-negative,
		\begin{align}
			\label{wec}
			\rho\ge 0\;,
		\end{align}
		and the density and pressures satisfy
		\begin{align}
			\label{wec1}
			\rho+p_r\ge 0 \quad \text{and} \quad \rho+p_t\ge 0 \;.
		\end{align}
		
		\item[(2)] Strong Energy Condition (SEC): The SEC requires that the matter fields satisfy the following inequalities
		\begin{align}
			\label{sec}
			(n-3)\rho + p_r + (n-2)p_t \ge 0 \quad \text{and} \quad \rho+p_r\ge 0,\quad \rho+p_t\ge 0\;.
		\end{align}
	\end{itemize}

	\section{Lower bounds on the photon sphere radius and shadow radius}\label{lower-bounds}
	
	\subsection{Photon spheres and the capture shadow}\label{capture-subsection}
	
	By spherical symmetry, we may choose coordinates such that a null geodesic lies in the equatorial plane, with
	\begin{align}
		\theta_1=\theta_2=\cdots=\theta_{n-3}=\frac{\pi}{2}\;,\qquad
		\theta_{n-2}=\phi\;,
	\end{align}
	where $\phi$ is the azimuthal coordinate. Let $E=\mu e^{-2\delta}\dot t$ and $L=r^2\dot\phi$ be the conserved energy and angular momentum, with $E>0$ and $L\ge0$, and define the impact parameter $b=L/E$. A dot denotes differentiation with respect to an affine parameter. The null condition gives
	\begin{align}
		\dot r^2=E^2e^{2\delta}\left[1-b^2U(r)\right],\qquad
		U(r)=\frac{\mu(r)e^{-2\delta(r)}}{r^2}\;.
		\label{U}
	\end{align}
	Using Eqs.~(\ref{mu'}) and (\ref{delta'}), one obtains
	\begin{align}
		U'(r)=-\frac{e^{-2\delta}}{r^3}\mathcal R(r),\qquad
		\mathcal R(r)=3-n+(n-1)\mu-\frac{16\pi r^2p_r}{n-2}\;.
		\label{R}
	\end{align}
	A circular null geodesic at $r=r_\gamma>r_H$ has $L>0$ and satisfies $\dot r=\ddot r=0$, which gives
	\begin{align}
		\label{circular-null-conditions}
		b_\gamma^2U(r_\gamma)=1,\qquad U'(r_\gamma)=0\;.
	\end{align}
	Thus the photon sphere equation is $\mathcal R(r_\gamma)=0$, and the corresponding impact parameter is $b_\gamma=1/\sqrt{U(r_\gamma)}$. Here $r_\gamma$ denotes the areal radius of an arbitrary exterior photon sphere. A photon sphere at a nondegenerate local maximum of $U$ is radially unstable, while one at a nondegenerate local minimum is radially stable~\cite{Cardoso:2008bp}. Degenerate stationary points require a separate stability analysis, but the radius bounds derived below do not assume nondegeneracy or a particular radial stability.
	
	We assume that photons follow metric null geodesics, the exterior matter is transparent, and the horizon absorbs incident photons. Since $U$ is continuous and positive outside the horizon and tends to zero at both the horizon and infinity, it attains a positive global maximum at some $r_*>r_H$. An incoming ray from infinity with $b^2\max U<1$ has no turning point and is captured, whereas a ray with $b^2\max U>1$ cannot cross the exterior potential barrier and reach the horizon. Thus the asymptotic shadow radius is given by~\cite{Broderick:2023shadow}
	\begin{align}
		\label{shadow-global}
		r_{\mathrm{sh}}=b_c=\left[\max_{r>r_H}U(r)\right]^{-1/2}=\frac{1}{\sqrt{U(r_*)}}\;.
	\end{align}
	Here $r_*$ denotes a photon sphere radius satisfying $U(r_*)=\max_{r>r_H}U(r)$ and $U'(r_*)=0$, and $r_{\mathrm{sh}}$ is the impact-parameter radius on an asymptotic observer's screen. For a spacetime with a single exterior photon sphere, $r_*=r_\gamma$. Eq.~(\ref{shadow-global}) remains valid for degenerate or multiple global maxima.
	
	\subsection{Lower bound on the photon sphere radius}\label{photon-section}
	
	The four-dimensional lower bound $r_\gamma\ge3r_H/2$ is known to require assumptions beyond the WEC. More precisely, Hod proved $r_{\gamma,\mathrm{in}}\ge3r_H/2$ for the innermost photon sphere of spherically symmetric black holes under the dominant energy condition, $T\le0$, and a monotonicity condition on the radial pressure~\cite{Hod:2020pim}, while Yang and L\"u formulated a sufficient auxiliary-function condition for the outermost photon sphere, noted that their proof also bounds the innermost one, and showed explicitly that the WEC alone does not guarantee the bound~\cite{Yang:2019zcn}. The same distinction applies to the mass-dependent upper bounds: Hod's theorem gives $r_{\gamma,\mathrm{in}}\le3M$ for the innermost photon sphere~\cite{Hod:2013jhd}, whereas Yang and L\"u established an upper bound for the outermost photon sphere under their stated assumptions~\cite{Yang:2019zcn}. Recently, Song et al.~\cite{Song:2026cci} generalized the photon sphere bounds to $n$-dimensional Einstein gravity. For the innermost photon sphere, they obtained the upper bound $r_\gamma\le[(n-1)M]^{1/(n-3)}$ under their pressure bounds $\rho\ge|p_r|,|p_t|$ and a non-positive trace condition, and the lower bound $r_\gamma\ge[(n-1)/2]^{1/(n-3)}r_H$ with the additional assumption that $|r^{n-1}p_r(r)|$ is non-increasing. Their lower-bound argument states the monotonicity condition outside the horizon, although only the interval between the horizon and the innermost photon sphere is needed for the proof. We first derive the same lower bound for an arbitrary exterior photon sphere under an auxiliary-function condition. We subsequently show that the WEC and pressure monotonicity up to the innermost photon sphere suffice to obtain the same bound for every exterior photon sphere, without requiring $T\le0$.
	
	Assume there exists a function $\Xi(r)$, continuous up to the horizon and differentiable in the exterior, such that~\cite{Yang:2019zcn}
	\begin{align}
		\label{Xi}
		-\rho(r)\le\Xi(r)\le p_r(r),\qquad [r^2\Xi(r)]'\ge0\;.
	\end{align}
	For a particular photon sphere, the auxiliary condition is needed only between the horizon and that photon sphere. Eq.~(\ref{Xi}) is imposed as an additional sufficient condition. It implies $\rho+p_r\ge0$; the proof below does not separately require $\rho\ge0$ or $\rho+p_t\ge0$. Thus the full WEC is not an additional assumption for this auxiliary-function proof.
	
	Let $r_\gamma>r_H$ denote the radius of an arbitrary exterior photon sphere. Write $k=n-3>0$ and $a=8\pi/(n-2)$. Define
	\begin{align}
		\mathcal M(r)=m(r)+\frac{a}{k+2}r^{k+2}p_r(r)\;.
	\end{align}
	Setting $\mathcal R(r_\gamma)=0$ in Eq.~(\ref{R}) and using Eq.~(\ref{mu(mr)}) gives
	\begin{align}
		\mathcal M(r_\gamma)=\frac{r_\gamma^k}{k+2}\;.
	\end{align}
	For $r_H\le x\le r\le r_\gamma$, monotonicity implies $x^2\Xi(x)\le r^2\Xi(r)$. Hence
	\begin{align}
		\label{massXi}
		m(r)&\ge\frac{r_H^k}{2}-a\int_{r_H}^r x^{k+1}\Xi(x)\D x\nonumber\\
		&\ge\frac{r_H^k}{2}-\frac{a}{k}r^2\Xi(r)(r^k-r_H^k)\;.
	\end{align}
	Using $p_r\ge\Xi$ in $\mathcal M$ yields
	\begin{align}
		\mathcal M(r)\ge\frac{r_H^k}{2}
		+\frac{a r^2\Xi(r)}{k}\left(r_H^k-\frac{2r^k}{k+2}\right)\;.
	\end{align}
	At a photon sphere this factorizes as
	\begin{align}
		\label{factor}
		\left[r_\gamma^k-\frac{k+2}{2}r_H^k\right]
		\left[1+\frac{2a}{k}r_\gamma^2\Xi(r_\gamma)\right]\ge0\;.
	\end{align}
	It remains to show strict positivity of the second factor on $(r_H,r_\gamma]$, where the auxiliary condition is assumed. By Eq.~(\ref{rho}) and the two-sided bound on $\Xi$, $\Xi(r_H)=-\rho(r_H)$ and
	\begin{align}
		F(r)\equiv1+\frac{2a}{k}r^2\Xi(r),\qquad F(r_H)\ge0\;.
	\end{align}
	Eq.~(\ref{Xi}) implies $F^\prime(r)=(2a/k)[r^2\Xi(r)]^\prime\ge0$ on $(r_H,r_\gamma)$, so $F(r)\ge0$ on $[r_H,r_\gamma]$. If $F(r)=0$ at a point in $(r_H,r_\gamma]$, Eq.~(\ref{massXi}) would imply $m(r)\ge r^k/2$, contrary to $\mu(r)>0$. Thus $F(r)>0$ throughout $(r_H,r_\gamma]$, including when the horizon is extremal. Dividing Eq.~(\ref{factor}) by $F(r_\gamma)$ and restoring $k=n-3$, we obtain
	\begin{align}
		\label{photon_lower_general}
		r_\gamma\ge\left(\frac{n-1}{2}\right)^{\frac{1}{n-3}}r_H\;.
	\end{align}
	Since $r_\gamma$ is arbitrary, Eq.~(\ref{photon_lower_general}) applies to any exterior photon sphere for which Eq.~(\ref{Xi}) holds on $[r_H,r_\gamma]$. If the auxiliary condition holds throughout the exterior, the bound applies to every exterior photon sphere, irrespective of its radial stability.
	For $n=4$, our result reduces to that of Yang and L\"u~\cite{Yang:2019zcn}.
	
	In particular, for a radial Maxwell field in $n$ dimensions, $p_r=-\rho=-D r^{-2(n-2)}$ with $D\ge0$. Choosing $\Xi=p_r$ gives
	\begin{align}
		(r^2\Xi)'=2(n-3)D r^{-2(n-3)-1}\ge0\;,
	\end{align}
	so this class satisfies the auxiliary condition. The vacuum also satisfies it with $\Xi=0$ and saturates Eq.~(\ref{photon_lower_general}).
	
	The pressure-monotonicity assumptions can be related directly to this proof. Suppose that an innermost exterior photon sphere exists. Since $U$ is positive outside the horizon and vanishes at the horizon, and $U'$ is continuous and has no zero before the first exterior stationary point, $U'$ has a constant positive sign on this interval. Thus $\mathcal R<0$ between the horizon and the innermost exterior photon sphere. From Eq.~(\ref{p'}),
	\begin{align}
		\label{pressure-weight-identity}
		(r^n p_r)'=\frac{r^{n-1}}{2\mu}\left[\mathcal R(\rho+p_r)+2\mu T\right]\le0
		\qquad (r_H<r<r_{\gamma,\mathrm{in}})\;,
	\end{align}
	where the inequality uses $\mathcal R<0$ on this interval, the WEC, and $T\le0$. Together with $p_r(r_H)=-\rho(r_H)\le0$, this gives $p_r\le0$ up to the innermost photon sphere. However, the trace condition can be dispensed with if $q(r)=|r^{n-1}p_r(r)|$ is non-increasing on $[r_H,r_{\gamma,\mathrm{in}}]$: continuity prevents $p_r$ from changing from a non-positive value to a positive value without passing through a zero, after which the non-negative, non-increasing function $q$ must remain zero up to $r_{\gamma,\mathrm{in}}$. Thus $p_r\le0$ on this interval follows directly from the horizon condition and pressure monotonicity. Since then $q=-r^{n-1}p_r$, the assumed monotonicity $q'\le0$ implies $(r^{n-1}p_r)'\ge0$, and
	\begin{align}
		\label{pressure-to-Xi}
		(r^2p_r)'=r^{3-n}(r^{n-1}p_r)'-(n-3)rp_r\ge0\;.
	\end{align}
	Thus $\Xi=p_r$ satisfies Eq.~(\ref{Xi}) up to $r_{\gamma,\mathrm{in}}$, and Eq.~(\ref{photon_lower_general}) gives the innermost lower bound under the WEC and pressure monotonicity on $[r_H,r_{\gamma,\mathrm{in}}]$. Since every exterior photon sphere satisfies $r_\gamma\ge r_{\gamma,\mathrm{in}}$, we obtain
	\begin{align}
		\label{photon_lower_from_in}
		r_\gamma\ge r_{\gamma,\mathrm{in}}\ge\left(\frac{n-1}{2}\right)^{\frac{1}{n-3}}r_H\;.
	\end{align}
	Neither $T\le0$ nor pressure monotonicity beyond $r_{\gamma,\mathrm{in}}$ is needed for this argument, as reflected in Table~\ref{tab:bounds}.

	\subsection{Lower bound on the shadow radius}\label{shadow-section}
	
	In this subsection we derive the lower bound on the shadow radius $r_{\mathrm{sh}}$ seen by an asymptotic observer. Using the radius $r_*$ defined in Eq.~(\ref{shadow-global}), we have
	\begin{align}
		\label{rsh2}
		r_{\mathrm{sh}}=\frac{r_*}{\sqrt{\mu(r_*) e^{-2\delta(r_*)}}} = \frac{1}{\sqrt{U(r_*)}} \;.
	\end{align}
	
	We now study the lower bound on the dimensionless ratio $r_{\mathrm{sh}}/r_H$ under $\rho\ge0$ and $\rho+p_r\ge0$, both implied by the WEC. Using Eq.~(\ref{mu(mr)}) in Eq.~(\ref{rsh2}), we have
	\begin{align}
		\label{rsh1}
		r_{\mathrm{sh}} = \frac{r_*}{\sqrt{1-\frac{2m(r_*)}{r_*^{n-3}}}}\,e^{\delta(r_*)} \;.
	\end{align}
	From $\rho+p_r\ge0$, which follows from the WEC, and Eq.~(\ref{delta'}), we have $\delta'(r)\le 0$. Since $\delta(\infty)=0$ by asymptotic flatness, it follows that $\delta(r)\ge 0$ for all finite $r\ge r_H$, giving $e^{\delta(r_*)}\ge 1$. Moreover, $\rho\ge 0$ implies $m(r)$ is non-decreasing (see Eq.~(\ref{m(r)})), hence $m(r_*)\ge m(r_H)=\frac12 r_H^{n-3}$. Therefore,
	\begin{align}
		\mu(r_*) = 1 - \frac{2m(r_*)}{r_*^{n-3}} \le 1 - \frac{r_H^{n-3}}{r_*^{n-3}} \;.
	\end{align}
	Combining these inequalities we obtain
	\begin{align}
		r_{\mathrm{sh}} \ge \frac{r_*}{\sqrt{1 - (r_H/r_*)^{n-3}}} \;.
	\end{align}
	Let $x = r_*/r_H > 1$. Then
	\begin{align}
		\frac{r_{\mathrm{sh}}}{r_H} \ge \frac{x}{\sqrt{1 - x^{-(n-3)}}} \equiv f(x)\;.
	\end{align}
	The function $f(x)$ is positive and tends to infinity as $x\to1^+$ and as $x\to\infty$. Its derivative is
	\begin{align}
		\label{fprime}
		f'(x)=\frac{1-\frac{n-1}{2}x^{-(n-3)}}{\left[1-x^{-(n-3)}\right]^{3/2}}\;.
	\end{align}
	For $x>1$, the denominator is positive and the numerator changes sign exactly once, from negative to positive, at $x^{n-3}=(n-1)/2$. Thus the unique global minimum occurs at
	\begin{align}
		x_{\min} = \left(\frac{n-1}{2}\right)^{\frac{1}{n-3}} \;.
	\end{align}
	Substituting back, we find
	\begin{align}
		f(x_{\min}) = \left(\frac{n-1}{2}\right)^{\frac{1}{n-3}} \sqrt{\frac{n-1}{n-3}} \;.
	\end{align}
	Thus we obtain the universal lower bound
	\begin{align}
		\label{shadow_lower_final}
		\frac{r_{\mathrm{sh}}}{r_H} \ge \left( \frac{n-1}{2} \right)^{\frac{1}{n-3}} \sqrt{\frac{n-1}{n-3}} \;.
	\end{align}
	The bound is saturated by the Schwarzschild--Tangherlini vacuum solution. The only matter inequalities used in the derivation are $\rho\ge0$ and $\rho+p_r\ge0$; the tangential inequality $\rho+p_t\ge0$ is not needed. Thus the full WEC is a sufficient hypothesis, while the two radial matter inequalities already suffice for this proof. In four dimensions, it reduces to
	\begin{align}
		\frac{r_{\rm sh}}{r_H}\geq \frac{3\sqrt{3}}{2},
	\end{align}
	in agreement with Ref.~\cite{Hod:2026jtr}.
	
	The same result can also be obtained directly from a pointwise comparison of the null potentials. Since $\rho\ge0$, Eq.~(\ref{m(r)}) implies
	\begin{align}
		m(r)\ge m(r_H)=\frac12 r_H^{n-3}\;,
	\end{align}
	and hence, using Eq.~(\ref{mu(mr)}),
	\begin{align}
		\mu(r)\le1-\left(\frac{r_H}{r}\right)^{n-3}\;.
	\end{align}
	Moreover, $\rho+p_r\ge0$ and Eq.~(\ref{delta'}) give $\delta'(r)\le0$. Together with $\delta(\infty)=0$, this yields $\delta(r)\ge0$ for $r>r_H$, and therefore $e^{-2\delta(r)}\le1$. Since $\mu(r)>0$ in the exterior, we obtain the pointwise bound
	\begin{align}
		\label{pointwise-lower}
		U(r)=\frac{\mu(r)e^{-2\delta(r)}}{r^2}
		\le\frac{1-(r_H/r)^{n-3}}{r^2}
		\qquad (r>r_H)\;.
	\end{align}
	Define
	\begin{align}
		U_H(r)\equiv\frac{1-(r_H/r)^{n-3}}{r^2}\;,
	\end{align}
	which is the null potential of the Schwarzschild--Tangherlini solution with the same horizon radius. Since Eq.~(\ref{pointwise-lower}) holds for every $r>r_H$, it follows immediately that
	\begin{align}
		\max_{r>r_H}U(r)\le\max_{r>r_H}U_H(r)\;.
	\end{align}
	This comparison does not require the two maxima to occur at the same radius. To evaluate the right-hand side, set $x=r/r_H>1$. Then
	\begin{align}
		U_H(r)=\frac{1}{r_H^2}\left[x^{-2}-x^{-(n-1)}\right]\;.
	\end{align}
	The derivative of the dimensionless factor is
	\begin{align}
		\frac{d}{dx}\left[x^{-2}-x^{-(n-1)}\right]
		=x^{-n}\left[(n-1)-2x^{n-3}\right]\;.
	\end{align}
	It changes sign exactly once, from positive to negative, at
	\begin{align}
		x^{n-3}=\frac{n-1}{2}\;.
	\end{align}
	Since $U_H$ vanishes as $x\to1^+$ and as $x\to\infty$, this stationary point is the unique global maximum for $x>1$. Therefore,
	\begin{align}
		\max_{r>r_H}U_H(r)
		=\frac{1}{r_H^2}\frac{n-3}{n-1}
		\left(\frac{2}{n-1}\right)^{\frac{2}{n-3}}\;.
	\end{align}
	Using Eq.~(\ref{shadow-global}), and noting that $y^{-1/2}$ is decreasing for $y>0$, we then find
	\begin{align}
		r_{\mathrm{sh}}
		&=\left[\max_{r>r_H}U(r)\right]^{-1/2}\\
		&\ge\left[\max_{r>r_H}U_H(r)\right]^{-1/2}\\
		&=r_H\left(\frac{n-1}{2}\right)^{\frac{1}{n-3}}
		\sqrt{\frac{n-1}{n-3}}\;,
	\end{align}
	which is precisely Eq.~(\ref{shadow_lower_final}). The comparison is pointwise throughout the exterior and therefore applies to configurations with any number of photon spheres. It constrains the height of the global potential barrier without imposing a lower bound on the radius at which that barrier occurs, and does not use Eq.~(\ref{Xi}) or pressure monotonicity.
	
	\subsection{Greater robustness of the shadow lower bound}\label{comparison-section}
	
	We now compare the two lower bounds within the class of spacetimes satisfying the WEC. The shadow lower bound in Eq.~(\ref{shadow_lower_final}) holds throughout this class under the stated geometric and photon-propagation assumptions. By contrast, the photon sphere lower-bound arguments in Sec.~\ref{photon-section} use an auxiliary-function condition or an additional pressure-monotonicity condition. To compare the two radii associated with the same photon sphere, we first use the radius $r_*$ attaining the global maximum of $U$. Eq.~(\ref{shadow-global}) gives
	\begin{align}
		\label{shadow-photon}
		r_{\mathrm{sh}}=\frac{r_*e^{\delta(r_*)}}{\sqrt{\mu(r_*)}}\;.
	\end{align}
	Here $r_*$ is the areal radius of a photon sphere determining the capture shadow, while $r_{\mathrm{sh}}$ is its critical impact parameter. For any exterior photon sphere, the global definition of the shadow implies
	\begin{align}
		\label{photon-impact-comparison}
		b_\gamma=\frac{1}{\sqrt{U(r_\gamma)}}\ge r_{\mathrm{sh}}\;,
	\end{align}
	with equality if and only if $U(r_\gamma)=\max_{r>r_H}U(r)$. Thus an arbitrary photon sphere's associated impact parameter is not necessarily the capture threshold. The two coincide for the photon sphere at $r_*$ and, in particular, for a spacetime with a single exterior photon sphere.
	
	The proof of Eq.~(\ref{shadow_lower_final}) requires only $\rho\ge0$ and $\rho+p_r\ge0$, both implied by the WEC, and does not use the photon sphere lower bound in Eq.~(\ref{photon_lower_general}). In particular, if $r_*$ falls below $[(n-1)/2]^{1/(n-3)}r_H$ while the WEC holds, the metric factors in Eq.~(\ref{shadow-photon}) necessarily keep $r_{\mathrm{sh}}$ above its lower bound. At fixed horizon radius, the non-negative exterior energy density reduces $\mu$ relative to the vacuum value, while $\rho+p_r\ge0$ gives $e^{-2\delta}\le1$. Both effects lower the null potential pointwise relative to the vacuum solution with the same horizon radius and hence increase, or leave unchanged, the critical capture impact parameter. This constrains the height of the global potential barrier, while its location can move inward without violating the shadow lower bound.
	
	Within the WEC class, the auxiliary-function proof of the photon sphere lower bound uses Eq.~(\ref{Xi}) on the interval from the horizon to the sphere under consideration. The alternative argument based on the innermost photon sphere requires pressure monotonicity only on $[r_H,r_{\gamma,\mathrm{in}}]$ and then gives the bound for every exterior photon sphere by ordering; the trace condition used in Ref.~\cite{Song:2026cci} is not needed for this argument. The shadow lower bound is independent of these additional hypotheses. In this sense, the lower bound on the shadow radius is more robust than that on the photon sphere radius in $n$-dimensional Einstein gravity. This comparison concerns the assumptions of the proofs within the WEC class; the auxiliary-function condition by itself does not imply the full WEC, and the sufficient conditions given here are not asserted to be necessary for the photon sphere bound. The following four-dimensional example shows an actual violation of the photon sphere lower-bound inequality while the shadow lower bound remains valid, even though both radii are associated with the same photon sphere.
	
	\subsection{A four-dimensional matter example}\label{matter-example}
	
	Consider the four-dimensional metric and anisotropic stress tensor of Ref.~\cite{Yang:2019zcn}. This is an Einstein-matter configuration specified at the stress-tensor level; no particular matter Lagrangian or dynamical stability of the supporting matter is assumed. As shown below, this spacetime has exactly one exterior photon sphere, which determines the capture shadow. In units $r_H=1$, take
	\begin{align}
		\label{example}
		\delta=0,\qquad
		\mu(r)=1-\frac1r-\frac{6}{\sqrt e\,r}
		+e^{-r/2}\left(2+\frac4r\right)\;.
	\end{align}
	Direct substitution into Einstein's equations gives
	\begin{align}
		\label{densityexample}
		\rho=-p_r=\frac{e^{-r/2}}{8\pi r},\qquad
		p_t=\rho\left(\frac r4-\frac12\right)\;.
	\end{align}
	Thus $\rho\ge0$, $\rho+p_r=0$, and $\rho+p_t=\rho(r/4+1/2)>0$, so the WEC is satisfied throughout the exterior region.
	
	There is no second horizon outside $r=1$ and the photon sphere equation is 
	\begin{align}
		\label{rootexample}
		P(r)\equiv3\left(1+\frac6{\sqrt e}\right)-2r
		-e^{-r/2}(r^2+6r+12)=0\;.
	\end{align}
	Here $P'= -2+e^{-r/2}(r^2/2+r)<0$: the positive term has its global maximum at $r=1+\sqrt5$ and that maximum is $(4+2\sqrt5)e^{-(1+\sqrt5)/2}\simeq1.68<2$. Also $P(1)=1-e^{-1/2}>0$, whereas $P(r)\to-\infty$ as $r\to\infty$. Since $U'(r)=P(r)/r^4$ for this metric, the unique exterior root is a nondegenerate global maximum of $U$. There is therefore a single exterior photon sphere, with $r_\gamma=r_*=r_{\gamma,\mathrm{in}}=r_{\gamma,\mathrm{out}}$. Solving Eq.~(\ref{rootexample}) gives
	\begin{align}
		r_\gamma&\simeq1.417<1.5,\\
		\mu(r_\gamma)&\simeq0.101,\\
		r_{\mathrm{sh}}&\simeq4.464>\frac{3\sqrt3}{2}\;.
	\end{align}
	Because $p_r=-\rho$, Eq.~(\ref{Xi}) forces $\Xi=p_r$. In this case
	\begin{align}
		(r^2\Xi)'=-\frac{e^{-r/2}}{8\pi}\left(1-\frac r2\right)<0
		\qquad (1<r<2)\;,
	\end{align}
	so the monotonicity condition in Eq.~(\ref{Xi}) is violated between the horizon and the photon sphere.
	
	The assumptions used for the lower bound on the innermost photon sphere in Sec.~\ref{photon-section} can also be checked explicitly. For this configuration,
	\begin{align}
		\label{example-pressure-monotonicity}
		|r^3p_r|=\frac{r^2e^{-r/2}}{8\pi},\qquad
		\frac{d}{dr}|r^3p_r|=\frac{r e^{-r/2}}{8\pi}\left(2-\frac r2\right)>0
		\quad (1<r<4)\;.
	\end{align}
	
	Thus the required non-increase fails already on $[r_H,r_{\gamma,\mathrm{in}}]$, excluding this configuration from the pressure-monotonicity argument and consistently allowing the photon sphere lower-bound inequality to be violated. Hence this configuration satisfies $r_*=r_\gamma$ and the WEC, while
	\begin{align}
	r_\gamma<\frac{3r_H}{2}\;,\qquad \frac{r_{\mathrm{sh}}}{r_H}>\frac{3\sqrt3}{2}\;.
	\end{align}
	It therefore provides an explicit example of the greater robustness of the shadow lower bound in four dimensions: the WEC preserves the lower bound on the critical impact parameter even though the lower bound on the areal radius of the same photon sphere fails.

	\section{Conclusion}\label{section6}
	
	In this work, we have derived lower bounds on the shadow radius and the photon sphere radius for static, spherically symmetric, asymptotically flat black holes in $n$-dimensional Einstein gravity ($n\ge4$), supported by anisotropic matter. By comparing the matter assumptions of the two proofs, we have shown the sense in which the lower bound on the shadow radius is more robust than that on the photon sphere radius.
	
	Defining the shadow radius by the critical impact parameter for null-geodesic capture, we obtained
	\begin{align}
		\frac{r_{\mathrm{sh}}}{r_H}\ge\left(\frac{n-1}{2}\right)^{\frac{1}{n-3}}\sqrt{\frac{n-1}{n-3}}\;.
	\end{align}
	This bound follows from the WEC and holds whether the effective potential has one or several maxima. Its derivation uses only $\rho\ge0$ and $\rho+p_r\ge0$, together with the stated geometric and photon-propagation assumptions.
	
	For the photon sphere radius, the auxiliary-function condition $-\rho\le\Xi\le p_r$ and $(r^2\Xi)'\ge0$ on $[r_H,r_\gamma]$ yields
	\begin{align}
		r_\gamma\ge\left(\frac{n-1}{2}\right)^{\frac{1}{n-3}}r_H\;.
	\end{align}
	The weight $r^2\Xi$ in the mass integral permits a direct extension of the four-dimensional proof to arbitrary dimension. When the auxiliary-function condition holds throughout the exterior, the bound applies to every exterior photon sphere. Alternatively, the WEC and non-increase of $|r^{n-1}p_r|$ up to the innermost photon sphere give the same bound for every exterior photon sphere by ordering, without requiring $T\le0$. Both lower bounds are saturated by the vacuum Schwarzschild--Tangherlini solution and reduce to the corresponding four-dimensional results.
	
	The shadow radius is the critical impact parameter of a photon sphere attaining the global maximum of $U$, whose areal radius is $r_*$. For a general exterior photon sphere, $b_\gamma\ge r_{\mathrm{sh}}$, with equality precisely when that sphere attains the global maximum. The shadow lower bound remains independent of the additional hypotheses used in the photon sphere lower-bound proofs. The four-dimensional example has a single exterior photon sphere, so that $r_*=r_\gamma$, and satisfies the WEC while violating both the auxiliary-function condition and the assumptions of the pressure-monotonicity argument. It obeys the shadow lower bound despite $r_\gamma<3r_H/2$. Thus the distinction between the sufficient matter assumptions is established in arbitrary dimension, while an actual violation of the photon sphere lower-bound inequality within the WEC class is demonstrated in four dimensions. The mass-dependent upper bounds are presented in Appendix~\ref{appendixA} under the WEC, the SEC, and suitable asymptotic decay assumptions.
	
	Extensions to rotating black holes and asymptotically (anti-)de Sitter spacetimes remain interesting directions for future work. In the rotating case, the photon sphere is replaced by a photon region, and it would be natural to ask whether the greater robustness found here for shadow bounds persists when the shadow is determined by a family of critical null geodesics rather than a single spherical photon orbit. For asymptotically (anti-)de Sitter spacetimes, the altered asymptotic structure may require a reformulation of the global-potential argument and of the relation between critical impact parameters and the shadow. Another natural direction is to investigate whether analogous bounds continue to hold in higher-curvature theories of gravity or under weaker energy conditions.
	
	\appendix
	\section{Upper bounds on the photon sphere radius and shadow radius}\label{appendixA}
	
	In this Appendix we derive upper bounds on the photon sphere radius and the shadow radius, which require the weak energy condition (WEC), the strong energy condition (SEC), and an asymptotic decay condition on the matter fields. These results are summarized in Table~\ref{tab:bounds}.

	\subsection{Upper bound on the photon sphere radius}
	
	To establish the upper bound, we impose the WEC and the SEC throughout the exterior, together with the asymptotic decay assumptions in Eq.~(\ref{decay_assumption}). We use the potential $U(r)$ and the photon sphere condition $\mathcal R(r_\gamma)=0$ introduced in Sec.~\ref{capture-subsection}, where $r_\gamma$ may be any exterior photon sphere radius. The bound applies to every exterior photon sphere, without an assumption on its radial stability. The WEC gives $0<\mu(r)\le1$ and $\delta(r)\ge0$ for $r>r_H$.
	Introduce the auxiliary function
	\begin{align}
		\label{G}
		G(r)=r^{n-1}U = r^{n-3}\mu e^{-2\delta}\;.
	\end{align}
	Taking the derivative of $G(r)$ and inserting Eqs.~(\ref{mu'}) and (\ref{delta'}) yields
	\begin{align}
		\label{G'}
		G'(r)=r^{n-4}e^{-2\delta}\biggl[(n-3)+\frac{16\pi r^2 p_r}{n-2}\biggr]\;.
	\end{align}
	Introduce a new auxiliary function $W(r)$ defined as
	\begin{align}
		\label{W}
		W(r)=e^{-2\delta}\biggl[1+\frac{16\pi r^2 p_r}{(n-2)(n-3)}\biggr]\;,
	\end{align}
	so that
	\begin{align}
		\label{G'andW}
		G'(r)=(n-3)r^{n-4}W(r)\;.
	\end{align}
	To prove the key inequality $W(r)\le1$ throughout the exterior, we analyze the sign of $p_r$ in two complementary cases, following the comparison strategy of Ref.~\cite{Yang:2019zcn}. Let $I_j^+$ denote a connected component of the open set $\{r>r_H:p_r(r)>0\}$. The remaining points satisfy $p_r\le0$.
	\begin{itemize}
		\item[(1)] For $r\in I_j^+$, differentiating $W(r)$ and using Eqs.~(\ref{delta'}) and (\ref{p'}) gives
		\begin{multline}
			\label{W'}
			W'(r)=e^{-2\delta}\biggl\{\frac{16\pi r(\rho+p_r)}{(n-2)(n-3)\mu}\biggl(\frac{n-3}{2}+\frac{8\pi r^2 p_r}{n-2}\biggr)
			+\frac{16\pi r}{(n-2)(n-3)}\biggl[\frac{n-3}{2}\rho+\frac{5-n}{2}p_r+(n-2)p_t\biggr]\biggr\}\;.
		\end{multline}
		Since $0<\mu\le1$ and $p_r>0$ on $I_j^+$, we obtain
		\begin{align}
			W'(r)\ge e^{-2\delta}\biggl[\frac{16\pi r(\rho+p_r)}{(n-2)(n-3)}\biggl(\frac{8\pi r^2 p_r}{n-2}\biggr)+\frac{16\pi r}{(n-2)(n-3)}\bigl[(n-3)\rho+p_r+(n-2)p_t\bigr]\biggr]\;.
		\end{align}
		The WEC and SEC therefore imply $W'(r)\ge0$ on $I_j^+$. This ensures that $W(r)$ is non-decreasing on each positive-pressure interval $I_j^+$. Consequently, on an interval with a finite right endpoint, continuity gives $p_r=0$ at that endpoint, so the limiting value satisfies $W=e^{-2\delta}\le1$. Monotonicity therefore gives $W\le1$ throughout the interval. If the interval extends to infinity, the falloff assumptions below give $W\to1$, and monotonicity again implies $W\le1$.
		
		\item[(2)] If $p_r\le0$, from Eq.~(\ref{W}) we obtain directly
		\begin{align}
			\label{W<1}
			W(r)\leq e^{-2\delta}\leq 1\;.
		\end{align}
	\end{itemize}
	Combining the two cases gives $W(r)\le1$ for all $r>r_H$.
	
	Now construct a vacuum potential
	\begin{align}
		\label{Uvac}
		U_{\mathrm{vac}}=\frac{1-\frac{2M}{r^{n-3}}}{r^2}\;,
	\end{align}
	and define the corresponding auxiliary function $G_{\mathrm{vac}}=r^{n-1}U_{\mathrm{vac}}$, whose derivative is
	\begin{align}
		\label{Gvac'}
		G_{\mathrm{vac}}' = (n-3)r^{n-4}\;.
	\end{align}
	When $W(r)\le 1$, the following relation holds
	\begin{align}
		\label{Gvac'>G}
		G_{\mathrm{vac}}' \geq G' = (n-3)r^{n-4}W(r)\;.
	\end{align}
	To compare $G$ and $G_{\mathrm{vac}}$, we consider their difference $H(r)=G_{\mathrm{vac}}(r)-G(r)$. The derivative satisfies $H'(r)\ge 0$. The monotonicity of $H(r)$ will fix its sign once the integration constant is determined. To determine this constant, we examine the asymptotic behavior of $H(r)$. The asymptotic decay assumptions imply
	\begin{align}
		\label{assume}
	\lim_{r\rightarrow\infty} r^{n-1}\rho(r) = 0\;,\quad\mathrm{and}\quad \lim_{r\rightarrow\infty} r^{n-1}p_r(r) = 0\;, 
	\end{align}
	where we impose the sufficient power-law bounds
	\begin{align}
		\label{decay_assumption}
		\rho(r) = \mathcal{O}\bigl(r^{-(n-1+\varepsilon_1)}\bigr),\qquad p_r(r) = \mathcal{O}\bigl(r^{-(n-1+\varepsilon_2)}\bigr)\quad\text{as } r\to\infty\;,
	\end{align}
	where $\varepsilon_1>0$ and $\varepsilon_2>0$. These pointwise falloff assumptions are stronger than the finite-mass requirement. They imply $\mathcal R(r)\to2$ as $r\to\infty$, so $U'(r)<0$ sufficiently far from the horizon. The bounds below use the photon sphere condition at an arbitrary exterior stationary radius.
	
	From Eq.~(\ref{delta'}) and the bounds on $\rho+p_r$, we have for large $r$
	\begin{align}
		\delta'(r) = -\frac{8\pi r (\rho+p_r)}{(n-2)\mu} = \mathcal{O}\bigl(r^{-(n-2+\varepsilon)}\bigr),\qquad \varepsilon = \min(\varepsilon_1,\varepsilon_2)>0\;.
	\end{align}
	Integrating from $r$ to $\infty$ gives
	\begin{align}
		\delta(r) = \int_r^\infty (-\delta'(x))\,dx = \mathcal{O}\bigl(r^{-(n-3+\varepsilon)}\bigr)\;.
	\end{align}
	Hence
	\begin{align}
		\lim_{r\to\infty} r^{n-3}\delta(r)=0,\qquad e^{-2\delta(r)} = 1 + \mathcal{O}\bigl(r^{-(n-3+\varepsilon)}\bigr)\;.
	\end{align}
	Using the asymptotic mass parameter $M$, Eq.~(\ref{m(r)}) can be written as
	\begin{align}
		m(r) = M - \frac{8\pi}{n-2}\int_r^\infty \chi^{n-2}\rho(\chi)d\chi\;.
	\end{align}
	With $\rho(\chi)=\mathcal{O}(\chi^{-(n-1+\varepsilon_1)})$, the integral is $\mathcal{O}(r^{-\varepsilon_1})$, so
	\begin{align}
		\mu(r) = 1 - \frac{2M}{r^{n-3}} + \mathcal{O}\bigl(r^{-(n-3+\varepsilon_1)}\bigr)\;.
	\end{align}
	Combining these, we find
	\begin{align}
		G(r) &= r^{n-3}\mu(r)e^{-2\delta(r)} \nonumber\\
		&= r^{n-3}\Bigl(1 - \frac{2M}{r^{n-3}} + \mathcal{O}(r^{-(n-3+\varepsilon_1)})\Bigr)\Bigl(1 + \mathcal{O}(r^{-(n-3+\varepsilon)})\Bigr) \nonumber\\
		&= r^{n-3} - 2M + \mathcal{O}\bigl(r^{-\varepsilon}\bigr)\;.
	\end{align}
	Therefore,
	\begin{align}
		H(r) = G_{\mathrm{vac}}(r)-G(r) = \mathcal{O}\bigl(r^{-\varepsilon}\bigr) \to 0 \quad\text{as } r\to\infty\;.
	\end{align}
	Because $H(r)$ is non-decreasing (since $H'(r)\ge0$) and tends to zero at infinity, it is non-positive for every $r>r_H$:
	\begin{align}
		\label{H<0}
		H(r)\le 0 \quad \Longrightarrow \quad G_{\mathrm{vac}}(r)\le G(r)\;.
	\end{align}
	From the definitions, this implies
	\begin{align}
		U_{\rm vac}(r)\leq U(r),\qquad r>r_H\;.
	\end{align}
	Thus the effective potential is bounded below by the Schwarzschild--Tangherlini potential with the same asymptotic mass parameter $M$.
	
	We now relate the position of the photon sphere to its vacuum counterpart. The photon sphere condition $U'(r_\gamma)=0$ is equivalent to $r_\gamma G'(r_\gamma)=(n-1)G(r_\gamma)$. Using $G'(r)\le G_{\mathrm{vac}}'(r)=(n-3)r^{n-4}$ and $G(r)\ge G_{\mathrm{vac}}(r)=r^{n-3}-2M$, we obtain
	\begin{align}
		(n-1)(r_\gamma^{n-3}-2M)&\le(n-1)G(r_\gamma)=r_\gamma G'(r_\gamma)\nonumber\\
		&\le r_\gamma G_{\mathrm{vac}}'(r_\gamma)=(n-3)r_\gamma^{n-3}\;.
	\end{align}
	Rearranging yields
	\begin{align}
		\label{upper r-gamma}
		r_\gamma \le \bigl[(n-1)M\bigr]^{\frac{1}{n-3}} \equiv r_{\gamma}^{\mathrm{vac}}\;.
	\end{align}
	This bound holds for every exterior photon sphere and is saturated by the vacuum Schwarzschild--Tangherlini black hole, whose photon sphere radius is $r_\gamma^{\mathrm{vac}}=[(n-1)M]^{1/(n-3)}$. For $n=4$, Eq.~(\ref{upper r-gamma}) reduces to
	\begin{align}
		r_\gamma\leq 3M,
	\end{align}
	which reproduces the four-dimensional photon sphere upper bound under the corresponding energy conditions~\cite{Yang:2019zcn}.
	
	This result has a simple geometric interpretation. For fixed asymptotic mass parameter \(M\) (equivalently, fixed physical ADM mass at a given dimension), no exterior photon sphere in the class of matter configurations considered here has an areal radius larger than the photon sphere of the Schwarzschild--Tangherlini spacetime. The inequality allows equality at the vacuum value.
	
	\subsection{Upper bound on the shadow radius}
	
	The upper bound on $r_{\mathrm{sh}}$ follows from the comparison $U\ge U_{\mathrm{vac}}$ for all $r>r_H$ and the global definition in Eq.~(\ref{shadow-global}). The vacuum potential has its maximum at
	\begin{align}
		r_\gamma^{\mathrm{vac}}=[(n-1)M]^{\frac{1}{n-3}}\;.
	\end{align}
	The WEC and Eq.~(\ref{m(r)}) give $2M\ge r_H^{n-3}$, so $r_\gamma^{\mathrm{vac}}>r_H$.  Consequently,
	\begin{align}
		\max_{r>r_H}U(r)\ge U(r_\gamma^{\mathrm{vac}})
		\ge U_{\mathrm{vac}}(r_\gamma^{\mathrm{vac}})
		=\frac{n-3}{n-1}\frac{1}{[(n-1)M]^{\frac{2}{n-3}}}\;.
	\end{align}
	Using Eq.~(\ref{shadow-global}) yields
	\begin{align}
		\label{shadow-upper}
		r_{\mathrm{sh}}\le\sqrt{\frac{n-1}{n-3}}\,[(n-1)M]^{\frac{1}{n-3}}\;.
	\end{align}
	The bound is saturated by the vacuum Schwarzschild--Tangherlini black hole and reduces to $r_{\mathrm{sh}}\le3\sqrt3 M$ in four dimensions~\cite{Yang:2019zcn}.

	\section*{Acknowledgments}
	This work is supported by the Research Start-up Funding of Chengdu University of Technology (Grant No. 10912-KYQD2022-09307).


\end{document}